\documentclass[aps,twocolumn,amsmath,amssymb,nofootinbib,superscriptaddress]{revtex4-2}
\usepackage{adjustbox}
\usepackage[utf8]{inputenc} 
\usepackage[colorlinks=true,
					linkcolor=magenta,
					anchorcolor=magenta,
					citecolor=magenta,
					urlcolor=magenta]{hyperref}        
\usepackage{url}            
\usepackage{booktabs}       
\usepackage{amsfonts}       
\usepackage{nicefrac}       
\usepackage{microtype}      
\usepackage{natbib}

\usepackage{bm}
\usepackage{bbm}
\usepackage{braket}
\usepackage{graphicx}
\usepackage{subfig}
\usepackage{float}
\usepackage{amsthm}
\usepackage{algorithmic}
\usepackage[linesnumbered,ruled]{algorithm2e}
\SetKwInOut{Parameter}{parameter}

\usepackage{mathtools}
\usepackage{nccmath}
\usepackage{color}
\usepackage{caption}
\usepackage{titlesec}

\usepackage{soul}
\titlespacing{\title}{0pc}{0.1pc}{0.3pc}
\titlespacing{\section}{0pc}{0.1pc}{0.3pc}

\begin{document}

\title{Control and mitigation of microwave crosstalk effect with superconducting qubits}

\author{Ruixia Wang}
\email{wangrx@baqis.ac.cn}
\affiliation{Beijing Academy of Quantum Information Sciences, Beijing 100193, China}%

\author{Peng Zhao}
\email{shangniguo@sina.com}
\affiliation{Beijing Academy of Quantum Information Sciences, Beijing 100193, China}%

\author{Yirong Jin}
\affiliation{Beijing Academy of Quantum Information Sciences, Beijing 100193, China}%

\author{Haifeng Yu}
\affiliation{Beijing Academy of Quantum Information Sciences, Beijing 100193, China}%

\date{\today}

\begin{abstract}
Improving gate performance is vital for scalable quantum computing. The universal quantum computing also requires the gate fidelity to reach a high level. For superconducting quantum processor, which operates in the microwave band, the single-qubit gates are usually realized with microwave driving. The crosstalk between microwave pulses is a non-negligible error source. In this article, we propose an error mitigation scheme to address this crosstalk issue for single-qubit gates. There are three steps in our method. First, by controlling the detuning between qubits, the microwave induced classical crosstalk error can be constrained within the computational subspace. Second, by applying the general decomposition procedure, arbitrary single-qubit gate can be decomposed as a sequence of $\sqrt{X}$ and virtual Z gates. Finally, by optimizing the parameters in virtual Z gates, the error constrained in the computational space can be corrected. Using our method, no additional compensation signals are needed, arbitrary single-qubit gate time will not be prolonged, and the circuit depth containing simultaneous single-qubit gates will also not increase. The simulation results show that, in specific regime of qubit-qubit detuning, the infidelities of simultaneous single-qubit gates can be as low as which without microwave crosstalk. 
\end{abstract}

\maketitle

\preprint{AIP/123-QED}

\bigskip
Quantum computing has developed from theoretical concept to experimental
realization, and it has been demonstrated that, quantum processor with
one- and two-qubit operations can construct the universal gate set
for quantum computing. However, the gate performance at present is insufficient
to achieve quantum advantage for practical applications \cite{preskill2018quantum}. Moreover conducting fault-tolerant quantum computing also requires the error of the quantum gate below the threshold
\cite{barends2014superconducting,blume2017demonstration,chow2012universal,fowler2012surface}. So that, reducing the one- and two-qubit gate errors is the key
goal at this stage. Crosstalk
is the leading source of the errors in gate operations, and it can occur in most
quantum systems and corrupt the quantum states when multiple quantum
gates are implemented simultaneously \cite{arute2019quantum,chen2022calibrated,sarovar2020detecting,zhao2022quantum}. This kind of error always violates
two key assumptions: spatial locality and independence of operations, and can be especially harmful to fault-tolerant quantum computing \cite{fowler2012surface,sarovar2020detecting}.

Crosstalk can be classified into two categories, one is classical
crosstalk, which is induced by the unintended classical electromagnetic couplings,
the other is quantum crosstalk arising from the residual quantum couplings
\cite{patterson2019calibration}. To address this issue, schemes for detecting, characterizing and analyzing the crosstalk
effects have been proposed \cite{gambetta2012characterization,rudinger2021experimental,huang2020alibaba,niu2021analyzing,ash2020experimental,dai2021calibration,abrams2019methods,rudinger2019probing,winick2021simulating,zajac2021spectator,sung2021realization,nuerbolati2022canceling,murali2020software,zhao2022spurious}. Hardware characterization protocols
are able to detect and identify various kinds of crosstalk errors
in multi-qubit processors by using Randomized Benchmarking \cite{gambetta2012characterization},
Gate Set Tomography \cite{rudinger2021experimental} or other methods \cite{ash2020experimental,dai2021calibration,abrams2019methods}. Crosstalk
mitigation schemes, such as applying a compensation pulse to cancel
the microwave crosstalk \cite{zajac2021spectator,sung2021realization,nuerbolati2022canceling}, or benchmarking the crosstalk induced error
first and then give an intelligent instruction through software techniques
\cite{murali2020software}, are also proposed. Characterizing and correcting all kinds of crosstalk
errors is an effective approach to reduce the gate
error.

For single-qubit gate operations, the average error for simultaneous
gates is larger than the isolated ones \cite{arute2019quantum,chen2022calibrated}.
One of the non-negligible noise for single-qubit gate is the classical
microwave crosstalk, which can be interpreted as the qubit feels the
unwanted microwave driving pulses implemented on the adjacent qubits. In order to mitigate such kind of error, the major method
is to cancel it actively by applying an out-of-phase compensation
signal \cite{sung2021realization,nuerbolati2022canceling}.

In this work, we present a systematic method for controlling and mitigating
the classical crosstalk effect for simultaneous single-qubit gates. This method is applicable for all kinds of systems which can use the virtual Z gate \cite{mckay2017efficient}. We take the system with frequency-tunable superconducting qubits as an example, our method can be applied as follows. First, by changing the qubit-qubit detuning, we can control the distribution of the microwave crosstalk induced errors, and constrain them in the computational space
instead of leading to leakage error. Second, as SU(2) gate can be parameterized as $U(\theta,\phi,\lambda)$ and decomposed as a sequence consisting of three Z gates and two $\sqrt{X}$ gates \cite{mckay2017efficient}, any single-qubit gate can be generated by tuning the parameters. Finally, by optimizing the parameters
in virtual Z gates in $U(\theta,\phi,\lambda)$, we can correct the single-qubit gate errors
induced by the microwave crosstalk during simultaneous gate operations. Note that, our proposed scheme is also compatible with fixed-frequency qubit architecture, which can be treated as a special case of the frequency-tunable one. In this case, we can skip the first step and apply the crosstalk mitigation method from the second step.

\begin{figure}[t!]
\centering{\includegraphics[width=85mm]{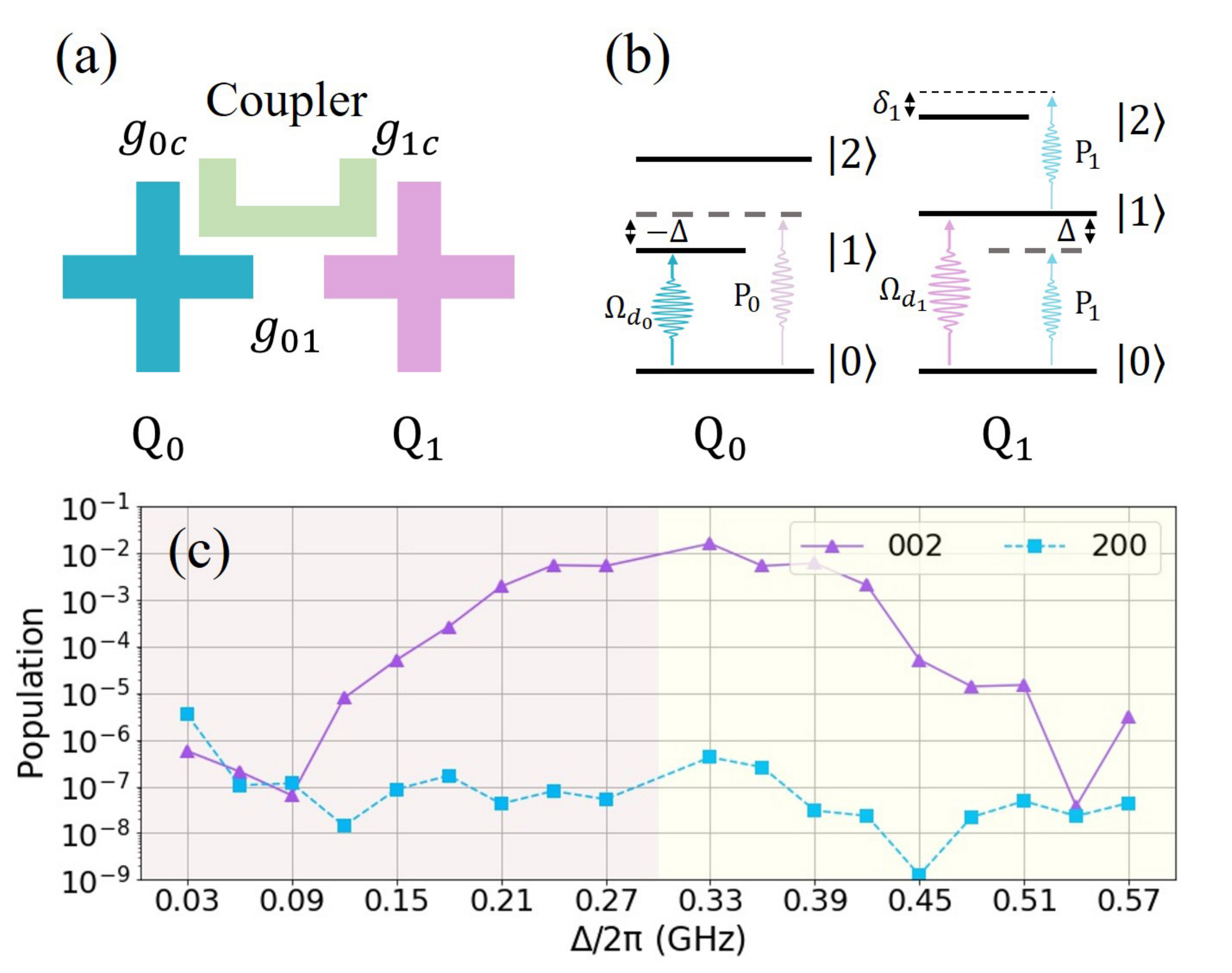}}
\caption{(a) Schematic diagram for the model we study. $\rm Q_{0}$ and $\rm Q_1$ are two transmon qubits. Coupler is an additional frequency-tunable qubit. $g_{01}$, $g_{0c}$ and $g_{1c}$ are the direct coupling strengths between $\rm Q_{0}$ and $\rm Q_1$, $\rm Q_{0}$ and coupler, $\rm Q_{1}$ and coupler, respectively. (b) The energy level for the two qubits. $\Delta=\omega_1-\omega_0$ is the qubit-qubit detuning. $\Omega_{d_0}$ and $\Omega_{d_1}$ are the strengths of the driving pulses on $\rm Q_{0}$ and $\rm Q_{1}$. $p_0$ ($p_1$) is cross driving strength from $\rm Q_{1(0)}$ to $Q_{0(1)}$. (c) The leakage to states $|002\rangle$ and $|200\rangle$ after two $\sqrt{X}$ gates for each qubit with the initial state of the system being prepared at the state $|101\rangle$. $\omega_0/2\pi = 5.34\, \rm GHz$, $\omega_1 = \omega_0+\Delta$, $\delta_{1}=\omega_{d_{0}}-\tilde{\omega}_{1}^{(21)}$, $\eta_0/2\pi=\eta_1/2\pi=-0.3\,  \rm GHz$, $\eta_c/2\pi = -0.1\, \rm GHz$, $g_{1c}/2\pi = g_{2c}/2\pi = 0.07\,  \rm GHz$, $g_{12}/2\pi = 0.005\, \rm GHz$, $p_0 = p_1 = 0.1$, and the $\sqrt{X}$ gate time is $12\,\rm ns$. The red shaded part is the straddling regime \cite{mundada2019suppression} and there is $\Delta<|\eta_{0(1)}|$, and in the yellow shaded part, there is $\Delta>|\eta_{0(1)}|$.} 
\label{f1}
\end{figure}

In the following, we give our analysis for the model with frequency-tunable qubits. The schematic diagram of our model is shown in FIG. \ref{f1}, where two of the frequency-tunable transmons act
as the qubits $\rm Q_{0,1}$ and the other one acts as a coupler $C$. The state can be denoted as $\rm |Q_{0}CQ_{1}\rangle$.
The Hamiltonian can be written as ($\hbar=1$):

\begin{eqnarray}
H&=&\sum_{l=0,1}(\omega_{l}a_{l}^{\dagger}a_{l}+\frac{\eta_{l}}{2}a_{l}^{\dagger}a_{l}^{\dagger}a_{l}a_{l})+\omega_{c}c^{\dagger}c+\frac{\eta_{c}}{2}c^{\dagger}c^{\dagger}cc\nonumber\\
&+&\sum_{l=0,1}g_{lc}(a_{l}^{\dagger}c+c^{\dagger}a_{l})+g_{01}(a_{0}^{\dagger}a_{1}+a_{1}^{\dagger}a_{0}),
\end{eqnarray}
where $a_l$ ($a_l^{\dagger}$) is the annihilation (creation) operator for $\rm Q_l$ with frequency $\omega_l$ and anharmonicity $\eta_l$, and $c$ ($c^{\dagger}$) is the annihilation (creation) operator for the coupler with frequency $\omega_c$ and anharmonicity $\eta_c$. $g_{lc}$ is the direct coupling strength between $\rm Q_l$ and the coupler $\rm C$, $g_{01}$ is the direct coupling strength between the two qubits. When operating the single-qubit gates, the qubit-qubit coupling is turned off by tuning the coupler. Thus single-qubit gates can be implemented without any residual inter-qubit couplings, such as XY- and ZZ-interactions (see supplementary material for ZZ-suppression point).

The single-qubit gate are implemented by applying a driving pulse to
the intended qubit. However, in the real quantum device, as the physical
isolation between the adjacent qubits is limited, there will always
be microwave crosstalk, such as due to the stray coupling between dedicated drive
lines, the direct coupling between one qubit, and the drive line of
others and the indirect coupling between one qubit and the others'
drive line  \cite{zhao2022spurious}. When operating the single-qubit gates, the accurate Hamiltonian without classical microwave crosstalk can be written as $H_{s}=H+H_{d}$, where

\begin{equation}
H_{d}=\sum_{l=0,1}[\Omega_{X_{l}}(t)\cos(\omega_{d_{l}}t)+\Omega_{Y_{l}}(t)\sin(\omega_{d_{l}}t)](a_{l}^{\dagger}+a_{l}),
\end{equation}
with $\Omega_{d_{l}}(t)\equiv\Omega_{X_{l}}(t)+i\Omega_{Y_{l}}(t)$. To suppress the leakage out of qubit space, we can use the derivative reduction by adiabatic gate (DRAG) pulse \cite{motzoi2009simple,chen2016measuring} to optimize the time-dependent driving pulse shape (see supplementary material for the optimization of the single-qubit gate). When the qubit $\rm Q_{0(1)}$ feels the driving pulse applied on $\rm Q_{1(0)}$, which is recorded as $\Omega_0 = p_{0}\Omega_{d_{1}}$
($\Omega_1 = p_{1}\Omega_{d_{0}}$), the crosstalk happens, where $p_{0}$ ($p_{1}$)
is constant and smaller than $1$. With the presence of microwave crosstalk, the Hamiltonian can be written as $H_{s}=H+H_{d}+H_{c}$, where

\begin{eqnarray}
H_{c}=\sum_{l=0,1}&&[p_l\Omega_{X_{1-l}}(t)\cos(\omega_{d_{1-l}}t+\tilde{\phi})+\nonumber\\
&&p_l\Omega_{Y_{1-l}}(t)\sin(\omega_{d_{1-l}}t+\tilde{\phi})](a_{l}^{\dagger}+a_{l}),
\end{eqnarray}
where $\tilde{\phi}$ is an additional phase. As shown in figure \ref{f1}(b), the
detuning between the qubit and the cross driving pulse is
$\Delta=\omega_{1}-\omega_{0}$ for $\rm Q_{1}$ ($-\Delta$ for $\rm Q_{0}$). Without loss of generality, we assume that $\omega_{1}>\omega_{0}$, giving rise to $\Delta>0$. When there is the detuning between the adjacent qubits, the classical microwave crosstalk induced error can be divided into two categories. 

The first kind is the error constrained within the computational space, such as the error induced by the AC Stark shift \cite{schuster2005ac} and bit-flip error under off-resonant drive \cite{malekakhlagh2022mitigating}.
With the AC Stark effect, the frequency of the qubit will change from $\tilde{\omega}_{0}$
($\tilde{\omega}_{1}$) to $\tilde{\omega}_{0}-\tilde{\Delta}_0$
($\tilde{\omega}_{1}+\tilde{\Delta}_{1}$), where $\tilde{\omega}_{l}$ represents the dressed frequency for qubit $\rm Q_l$ in the effective Hamiltonian (see supplementary material for the effective Hamiltonian), $\tilde{\Delta}_{l}=|p_{l}\Omega_{d_{1-l}}|^{2}/2\Delta$ ($l=0,1$).
Such a frequency shift can lead to an error as it induces the additional
terms $-\tilde{\Delta}_{0}a^{\dagger}_{0}a_{0}$ and $+\tilde{\Delta}_{1}a^{\dagger}_{1}a_{1}$
into the Hamiltonian. The bit-flip error is the cross-driving pulse induced $|0\rangle \leftrightarrow |1\rangle$ transition. Similar as the off-resonant driven Rabi oscilation, the bit-flip error rate is proportional to $p_l^2\Omega^2_{d_{1-l}}/(p_l^2\Omega^2_{d_{1-l}}+\Delta^2)$ for qubit $\rm Q_l$. Then for the first kind of error, it will become smaller as the qubit-qubit detuning $\Delta$ becomes larger. The second kind of the classical
crosstalk error is leakage error, which is caused by the unwanted
driving between the states $|1\rangle$ and $|2\rangle$. In FIG. \ref{f1}(b), the resonant frequency between the energy levels $|1\rangle$
and $|2\rangle$ is denoted as $\tilde{\omega}_{0}^{(12)}$ ($\tilde{\omega}_{1}^{(12)}$)
for $\rm Q_{0(1)}$. The detuning between the crosstalk driving pulse and
$\tilde{\omega}_l^{(12)}$ is $\delta_{0}=\omega_{d_{1}}-\tilde{\omega}_{0}^{(12)}$
($\delta_{1}=\omega_{d_{0}}-\tilde{\omega}_{1}^{(12)}$) for $\rm Q_{0(1)}$. When $\delta_l=0$, the leakage error for qubit $l$ reaches the maximum value. Because we have assumed that, $\omega_{1}>\omega_{0}$ and the anharmonicities
of the two qubits are the same, then there is $|\delta_{0}|>|\delta_{1}|$, so that, with the same driving and cross-driving strengths for the two qubits, the leakage error for $\rm Q_1$ will be larger than which for $\rm Q_0$, and when the value of the detuning $\Delta$ approaches $|\eta_{1}|$, the leakage error for $\rm Q_1$ will approaches the maximum.

FIG. \ref{f1}(c) shows the numerical results for the probabilities of the leakage error with different qubit-qubit detuning when operating the simultaneous single-qubit gates. Here the gate times of the two $\sqrt{X}$ gates are the same. As $|\delta_0|$ is larger than $|\delta_1|$, the leakage error for $\rm Q_0$ is smaller compared with $\rm Q_1$ for most values of $\Delta$. The leakage error on $\rm Q_1$ increases as $\Delta/ 2\pi$ increases from 0 to $0.3\, \rm GHz$, and at around $\Delta/ 2\pi\ = 0.3\, \rm GHz$, which equals the value of $|\eta_1/2\pi|$, the leakage error for $\rm Q_1$ reaches the maximum. Then, as the detuning continues to increase, the leakage error decreases. This tendency is consistent with our conclusion in the above. The small anomalies in the regime of $\Delta/2\pi <0.09 \, \rm GHz$ and $\Delta/2\pi >0.54 \, \rm GHz$ may come from the quantum crosstalk effect.

\begin{figure}[h]
\centering{\includegraphics[width=80mm]{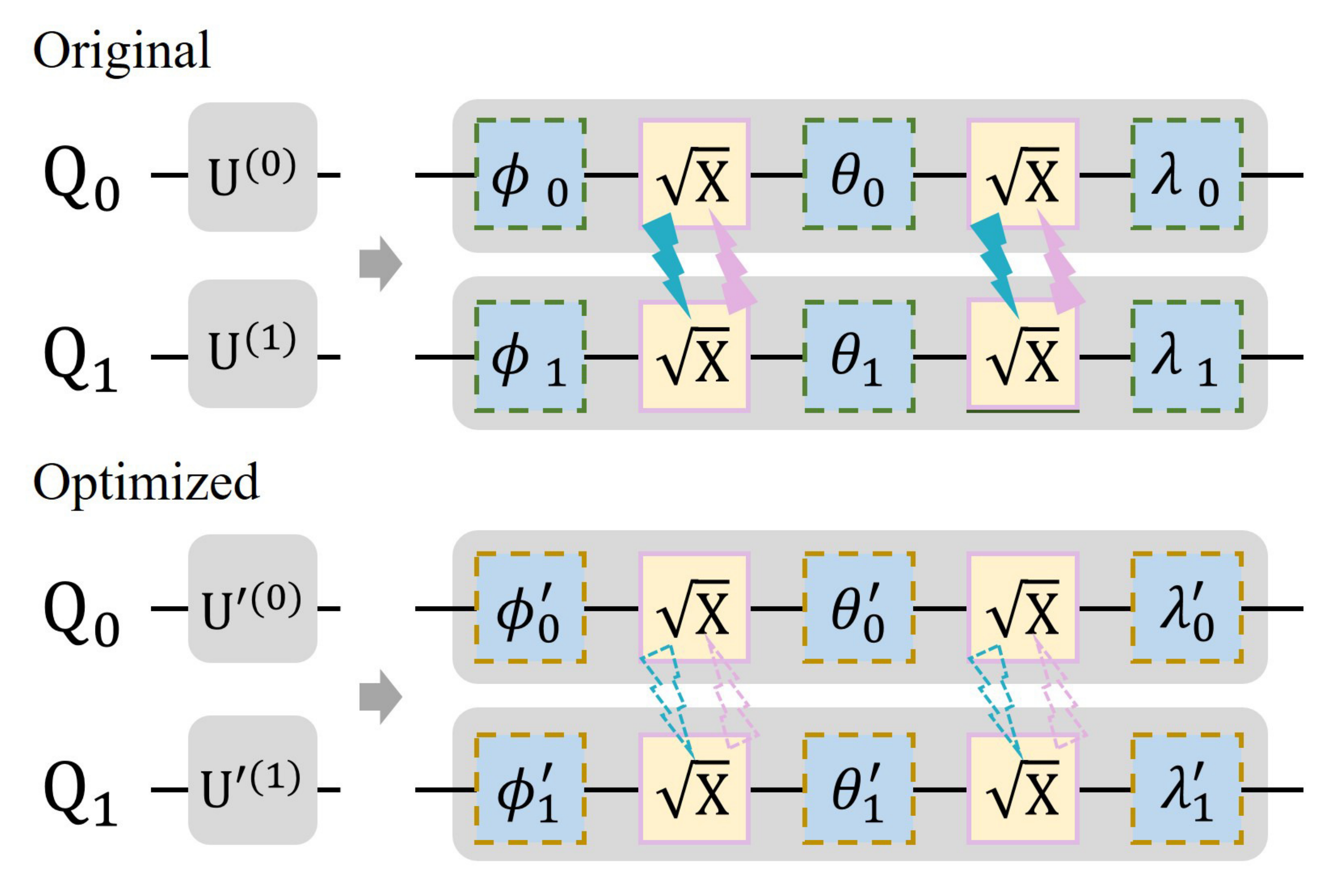}}
\caption{The original circuit is our target circuit, containing two arbitrary simultaneous single-qubit gates, which can be realized with three Z rotations and two $\sqrt{X}$ gates. The Z rotations can be operated using the virtual Z gate, then the crosstalk error happened only with the simultaneous $\sqrt{X}$ gates. The solid lightning symbols are used to indicate the strong microwave crosstalk. The optimized circuit is the one with the optimized parameters $\{\theta'_l, \phi'_l, \lambda'_l\}$ ($l=0,1$). Although the crosstalk still exists in the optimized circuit, the error constrained in the single-qubit computational space has been corrected with the optimized parameters. The hollow lightning symbols are used to indicate the mitigated microwave crosstalk effect.}
\label{f2}
\end{figure}

\begin{figure*}[ht!]
\centering{\includegraphics[width=160mm]{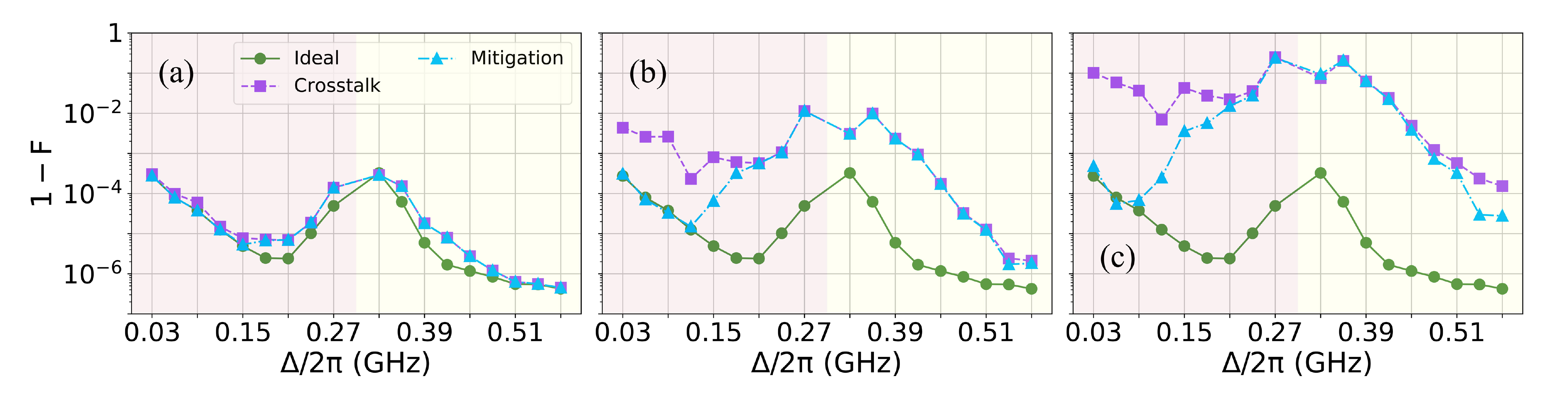}}
\caption{The infidelities of the circuit shown in FIG. \ref{f2}.  `Ideal' means there is no added classical microwave crosstalk. `Crosstalk' means that, we add the classical microwave crosstalk on the qubits when implementing the simultaneous $\sqrt{X}$ gates, the strengths of the crosstalk are $p_0=p_1=0.01$ for (a), $p_0=p_1=0.1$ for (b) and $p_0=p_1=0.5$ for (c). `Mitigation' represents the result after applying crosstalk mitigation scheme to the circuits with additional crosstalk. In our model, the systematic parameters are set as $\omega_0/2\pi=5.34 \, \rm GHz$, $\omega_1=\omega_0+\Delta$, $\eta_0/2\pi=\eta_1/2\pi=-0.3\,  \rm GHz$, $\eta_c/2\pi = -0.1\, \rm GHz$, $g_{1c}/2\pi = g_{2c}/2\pi = 0.07\,  \rm GHz$, $g_{12}/2\pi = 0.005\, \rm GHz$, $\theta_{0}/2\pi=0.704$, $\phi_{0}/2\pi=0.277$, $\lambda_{0}/2\pi=0.020$, $\theta_{1}/2\pi=0.987$, $\phi_{1}/2\pi=0.790$, $\lambda_{1}/2\pi=0.560$, and the $\sqrt{X}$ gate time is $12\,\rm ns$. The frequency of the coupler is at the ZZ-suppression point. The red shaded part is the straddling \cite{mundada2019suppression} regime and there is $\Delta<|\eta_{0(1)}|$, and in the yellow shaded part, there is $\Delta>|\eta_{0(1)}|$.}
\label{f3}
\end{figure*}

Then we would like to systematically give our protocol about control and mitigation of the crosstalk effect.
(1) Controlling the classical microwave crosstalk effect and constraining the error within the single-qubit computational space. From the above analysis, we know that by changing the qubit-qubit detuning, we can control the distribution of the two kinds of microwave crosstalk induced errors. And the error constrained in the single-qubit computational space can be corrected by single-qubit gates. While, the leakage error, which doesn't belong to the computational space, cannot be corrected only by single-qubit gate operations. So that, If we want to correct the error in gate level, we have to minimize the leakage error. As shown in FIG. \ref{f1}(c), in the first step, we can choose a regime, the straddling regime or out of the straddling regime, and then adjust the detuning to the point with the small leakage error. (2) Choosing the parameterized $U(\theta,\phi,\lambda)$ gate as the generalized decomposition for single-qubit gate. Any single-qubit gate can be realized by three rotations along Z axis and two rotations along X axis \cite{mckay2017efficient} as:

\begin{equation}
U(\theta,\phi,\lambda) = Z_{\phi-\pi/2}X_{\pi/2}Z_{\pi-\theta}X_{\pi/2}Z_{\lambda-\pi/2}.
\end{equation}
With most of the present quantum processors, we can apply a virtual Z gate with zero-duration instead of a real Z gate by changing the flux of the qubit. On these quantum processors, we can use the $U(\theta,\phi,\lambda)$ gate to generate arbitrary single-qubit gate in any quantum circuit and such a single-qubit gate only costs the time equal to two $\sqrt{X}$ gates. (3) Mitigating the error with the parameterized $U(\theta,\phi,\lambda)$ gates. Because the $U(\theta,\phi,\lambda)$ gate can generate arbitrary single-qubit gate, it can be used to correct the error constrained in the single-qubit computational space. As shown in FIG. \ref{f2}, in the original circuit, two single-qubit gates $U^{(l)}$ ($l=0,1$) are operated simultaneously, $\{\theta_l, \phi_l, \lambda_l\}$ ($l=0,1$) represent the parameters for the virtual Z gates.  When applying two $\sqrt{X}$ gates to the adjacent qubits simultaneously, because of the classical microwave crosstalk, the fidelity will decrease. To mitigate such error, we can optimize the parameters $\{\theta_l, \phi_l, \lambda_l\}$. 
After the optimization, the parameters are updated to $\{\theta'_{l}, \phi'_{l}, \lambda'_{l}\}$. As the parameters in $U(\theta,\phi,\lambda)$ are only related to the virtual Z gates, the optimization of them will not induce any new error or change the running time of the quantum circuit.

Next, we would like to give the numerical results for applying our scheme to mitigate the microwave crosstalk induced errors during simultaneous single-qubit gates.
The simultaneous single-qubit gates fidelity is defined as $F = \frac{\rm Tr(U_{imp}^{\dagger}U_{imp})+|\rm Tr(U^{\dagger}U_{imp})|^2}{d(d+1)}$ \cite{pedersen2007fidelity},
where $U$ and $U_{imp}$ represent the target gate operation and the implemented gate operation. $U=U^{(0)}\otimes U^{(1)}$ and $d$ is the system dimension.
We calculated the infidelities $1-F$ of the simultaneous single-qubit gates between two adjacent qubits with different values of detuning $\Delta$. As shown in FIG. \ref{f3},
in the straddling regime, we can see that, when the qubit-qubit detuning is relatively small, the infidelities with our error mitigation method are almost coincided with the ideal results, which implies that, in this case, the leakage error is much smaller than the error constrained in the computational space. So that, we can correct such error with the paremeterized $U(\theta,\phi,\lambda)$ gate. When the detuning $\Delta$ approaches the value of $|\eta_{0(1)}|$, the mitigated infidelity reachs the maximum value and is as high as the one without error mitigation, which means that, in this area, the leakage error is much larger than the error constrained in the computational space. Out of the straddling regime, as the detuning increases, both of the two kinds of classical microwave crosstalk induced errors decrease. The accuracy of the gate operations depends on a variety of factors, such as the classical microwave crosstalk induced error or the quantum crosstalk induced error. So that, to ensusre higher-fidelity single-qubit gates, we still have to avoid the frequency collision conditions, such as $\omega_0=\omega_1$ or $\omega_0=\omega_1+\eta_1$. As shown in FIG. \ref{f3}(c), the value of the optimal detuning with the minimum infidelity when applying our method may be smaller than which without our method, and as the strength of the crosstalk increases, the optimal value of $\Delta$ will decrease, which may provide the convenience for solving the crowded spectrum problem (similar results exist in FIG. 2(b) and (c) in the supplementary material).

Theoretically, this method is applicable to all strengths of the crosstalk. However, when the crosstalk is very weak, as shown in FIG. \ref{f3}(a), where $p_0=p_1=0.01$, i.e., the crosstalk strength is $-40\,\rm dB$, the error with the original circuit is much small, the mitigation effect is not obvious and there is no need to use this method. In FIG. \ref{f3}(b), the strength of the crosstalk is $-20\,\rm dB$, this strength is of general experimental interest \cite{gong2021quantum,ren2022experimental}. And in the worse case, the crosstalk strength may be as high as $-6\,\rm dB$ to $-10\,\rm dB$ in the experiment \cite{martinis2014ucsb}, just as shown in FIG. \ref{f3}(c), which is $-6\,\rm dB$. For the last two cases, our method shows good performance and significant reductions with the gate infidelities.

In this work, we proposed an error mitigation scheme for classical microwave crosstalk induced error by applying the advantage of the parameterized single-qubit $U(\theta,\phi,\lambda)$ gate. With proper detuning between the qubits, and optimizing the parameters $\{\theta, \phi, \lambda\}$, the significant reductions of the infidelities for simultaneous single-qubit gates can be realized, the value of the infidelities can be reduced to which without classical microwave crosstalk. Compared with other existing error mitigation methods, there are mainly three advantages. The first one is that, compared with the actively cancelling with the compensation pulse method \cite{sung2021realization,nuerbolati2022canceling}, our method can be done without applying additional compensation signal. The second one is that, we need not change the sequence of the gates or implement the simultaneous single-qubit gates in series, so that, we will not prolong the total operation time \cite{murali2020software}. The third one is that, using this method, we can get a more accurate circuit in the real experiment, so that, we can obtain the whole information of the final quantum states with high fidelities, instead of certain expectation values \cite{2017Error}.

There is also a limitation for our method. In this scheme, we only focus on the single-qubit gates. If we want to extend it to the two-qubit gates, we have to consider two cases. The first one is that, if the error is still constrained in the single-qubit computational space, it can be mitigate with our method  (see supplementary material for an example in this case). The second case is that, the error is not in the single-qubit computational space, but they are constrained in the two-qubit computational space. With our theory, such kind of error can be mitigated with the universal decomposition of two-qubit gate \cite{Vidal2004Universal}. However, there is no advantage like the virtual Z gate with the universal decomposition of the two-qubit gate. Arbitrary two-qubit gate can be decomposed into three two-qubit gates and several single-qubit gates, it will prolong the original circuit and consume more time. Then how to mitigate two-qubit gate error without additional time consuming is the key to the problem, and it is also an open question.

\bigskip

See the supplementary material for more detailed information. First, we give an anlysis for the ZZ-suppression point. Second, we talk about the optimization for the single-qubit gate. Third, we provide a supplemental numerical example to support our conclusion. Finally, we give an example for the error mitigation with two-qubit gate.

\bigskip

This work was supported by the Beijing Natural Science Foundation (Grant No. Z190012), the National Natural Science Foundation of China (Grants No. 11890704, No. 12004042, No. 12104055, No. 12104056, No. 12004206),
and the Key-Area Research and Development Program of Guang Dong Province (Grant No. 2018B030326001).

\bibliographystyle{apsrev4-2}
\bibliography{mainref}

\setcounter{equation}{0}
\setcounter{figure}{0}
\appendix

\renewcommand*{\thefigure}{S\arabic{figure}}
\renewcommand*{\theequation}{S\arabic{equation}}


\clearpage 
\bigskip

\section{The ZZ-suppression point}

In the main text, we have given the Hamiltonian for the system $\rm |Q_{0}CQ_{1}\rangle$ as ($\hbar=1$):

\begin{eqnarray}
H&=&\sum_{l=0,1}(\omega_{l}a_{l}^{\dagger}a_{l}+\frac{\eta_{l}}{2}a_{l}^{\dagger}a_{l}^{\dagger}a_{l}a_{l})+\omega_{c}c^{\dagger}c+\frac{\eta_{c}}{2}c^{\dagger}c^{\dagger}cc\nonumber\\
&+&\sum_{l=0,1}g_{lc}(a_{l}^{\dagger}c+c^{\dagger}a_{l})+g_{01}(a_{0}^{\dagger}a_{1}+a_{1}^{\dagger}a_{0}),
\end{eqnarray}

By applying the Schrieffer-Wolff transformation, the coupler is decoupled and the effective Hamiltonian
for the two qubits $\rm |Q_0Q_1\rangle$ can be approximated as:

\begin{equation}
H_{eff}=\tilde{\omega}_{0}\frac{ZI}{2}+\tilde{\omega}_{1}\frac{IZ}{2}+J\frac{XX+YY}{2}+\zeta\frac{ZZ}{4},
\end{equation}
where $X$, $Y$, $Z$ are the Pauli operators and $I$ is identity
operator. $\tilde{\omega}_{0(1)}$ is the dressed frequency for qubit
$\rm Q_{0(1)}$. The approximated XY coupling strength is $J=g_{01}+g_{0c}g_{1c}/\Delta_{12}$ \cite{yan2018tunable},
$1/\Delta_{12}=(1/\Delta_{1}+1/\Delta_{2})/2$  and
the ZZ coupling strength is defined as $\zeta=(E_{101}-E_{100})-(E_{001}-E_{000})$, where $E_{ijk}$ denotes the system's eigenenery for eigenstate $|ijk\rangle$. Usually, the system is operated in the dispersive and straddling regime. In the dispersive regime, there is $|\omega_{0(1)}-\omega_{c}|\gg g_{0c(1c)}$ \cite{blais2007quantum,mckay2016universal},
and in the straddling regime, there is $|\omega_{0}-\omega_{1}|<|\eta_{0(1)}|$
\cite{mundada2019suppression}. 

\begin{figure}[h]
\centering{\includegraphics[width=85mm]{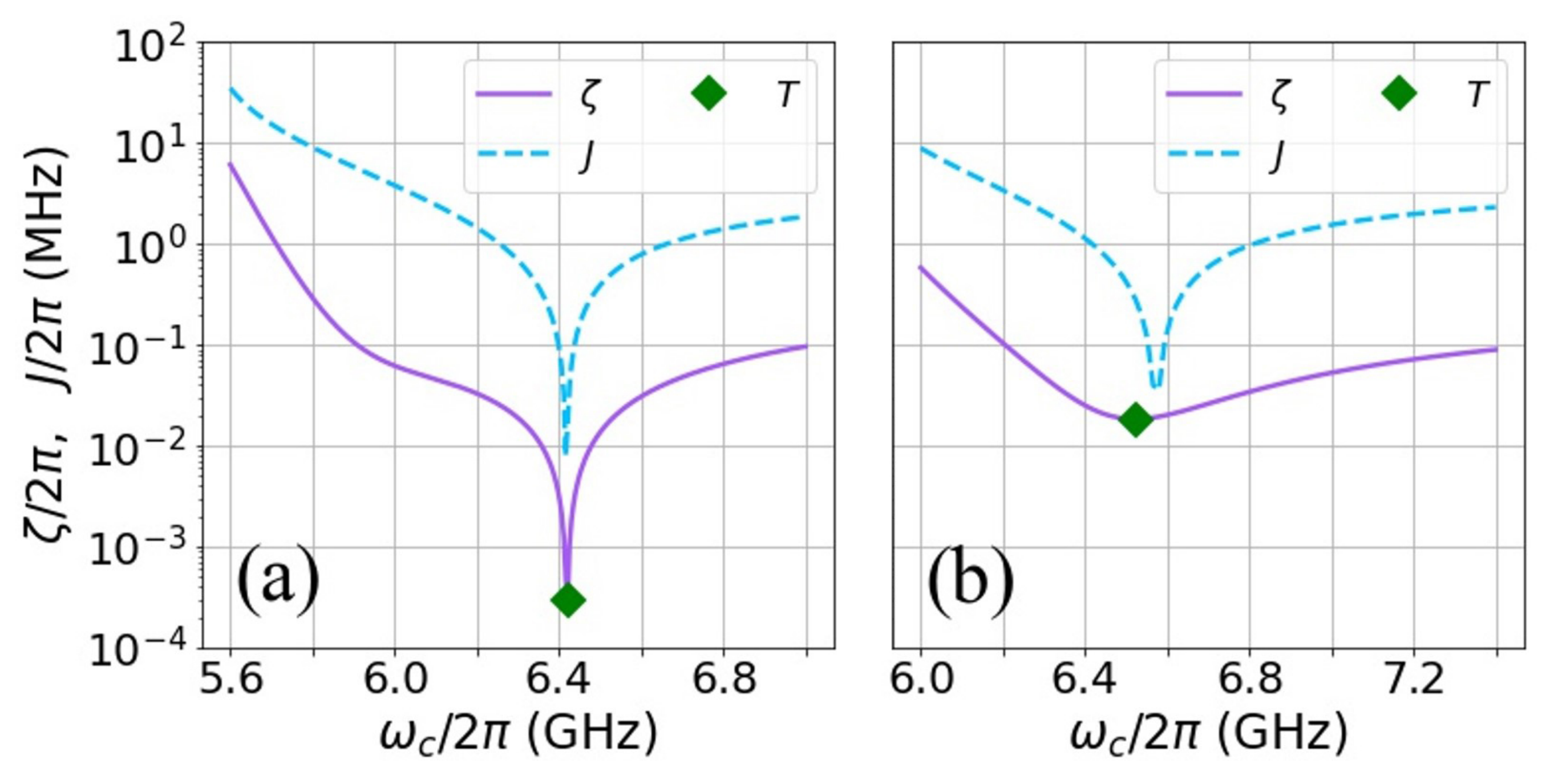}}
\caption{The XY and ZZ coupling strengths $J$ and $\zeta$ as the frequency of the coupler changes. (a) In this model, the qubits work in the straddling regime, $\omega_0/2\pi=5.34\, \rm GHz$, $\omega_1/2\pi=5.52 \, \rm GHz$, $\eta_0/2\pi=\eta_1/2\pi=-0.3\, \rm GHz$, $\eta_c/2\pi = -0.1 \, \rm GHz$. The direct coupling strengths are $g_{1c}/2\pi=g_{2c}/2\pi = 0.07\, \rm GHz$ and $g_{12}/2\pi=0.005\, \rm GHz$. (b) In this model, the qubits work out of the straddling regime, $\omega_0/2\pi=5.34\, \rm GHz$, $\omega_1/2\pi=5.76 \, \rm GHz$, $\eta_0/2\pi=\eta_1/2\pi=-0.3\, \rm GHz$, $\eta_c/2\pi = -0.1 \, \rm GHz$. The direct coupling strengths are $g_{1c}/2\pi=g_{2c}/2\pi = 0.07\, \rm GHz$ and $g_{12}/2\pi=0.005\, \rm GHz$.} 
\label{s1}
\end{figure}

When implementing single-qubit gate, we would like to trun off the ZZ interaction between the adjacent qubits. In the straddling regime, we can realize $\zeta=0$ by tuning the frequency of the coupler, as shown in FIG. \ref{s1}(a) with the point $T$. But out of the straddling regime, the numerical results show that, we can not achieve the point with $\zeta = 0$, which is shown in FIG. \ref{s1}(b). 

So that, if the frequencies of the qubits are in the straddling regime, we choose the ZZ-free point as the system's idle point, then
the single-qubit gates can be implemented without the residual ZZ
interaction, as shown in FIG. \ref{s1}(a) with the point $T$. And if the frequencies of the qubits are not in the straddling regime, we choose the point with the minimum ZZ interaction as the system's idle point, as the point $T$ shown in FIG. \ref{s1}(b). For these two cases, we uniformly refer to the $T$ point as ZZ-suppression point.

At the ZZ-suppression point, the effective
coupling strength $J/2\pi$ between them is about $1\,\rm MHz$ or smaller. If the detuning between the adjacent qubits $\Delta/2\pi$ is larger than $10\, \rm MHz$,  similar as the off-resonant driven Rabi oscillation, the swap error induced by XY coupling is proportional to $\frac{4J^{2}}{4J^{2}+\Delta^{2}}$, which is much small and can be neglected.

\begin{figure*}[ht!]
\centering{\includegraphics[width=170mm]{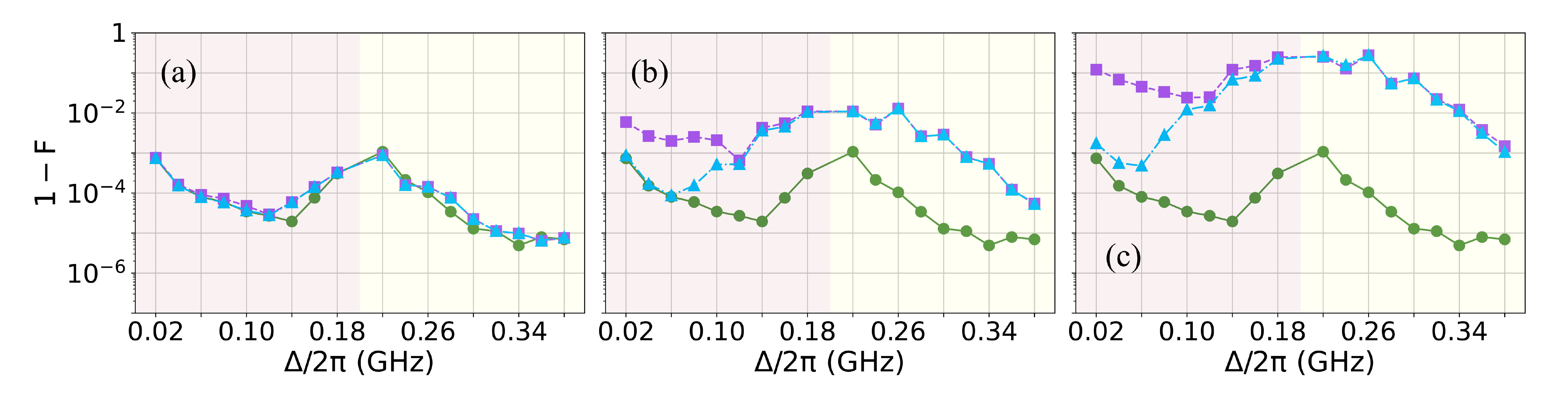}}
\caption{The infidelities of the circuit shown in FIG. 2 in the main text. `Ideal' means there is no added classical microwave crosstalk. `Crosstalk' means that, we add the classical microwave crosstalk on the qubits when implementing the simultaneous $\sqrt{X}$ gates, the strength of the crosstalk are $p_0=p_1=0.01$ for (a), $p_0=p_1=0.1$ for (b) and $p_0=p_1=0.5$ for (c). `Mitigation' represents the result after applying crosstalk mitigation scheme to the circuits with additional crosstalk. In our model, the system parameters are set as $\omega_0/2\pi=5.34 \, \rm GHz$, $\omega_1=\omega_0+\Delta$, $\eta_0/2\pi=\eta_1/2\pi=-0.2\,  \rm GHz$, $\eta_c/2\pi = -0.1\, \rm GHz$, $g_{1c}/2\pi = g_{2c}/2\pi = 0.07\,  \rm GHz$, $g_{12}/2\pi = 0.005\, \rm GHz$, $\theta_{0}/2\pi=0.704$, $\phi_{0}/2\pi=0.277$, $\lambda_{0}/2\pi=0.020$, $\theta_{1}/2\pi=0.987$, $\phi_{1}/2\pi=0.790$, $\lambda_{1}/2\pi=0.560$, and the $\sqrt{X}$ gate time is $12\,\rm ns$.  The frequency of the coupler is at the ZZ-free point in the straddling regime \cite{mundada2019suppression} or the minimum-ZZ-interaction point out of the straddling regime. The red shaded part is straddling regime, there is $\Delta<|\eta_{0(1)}|$, and in the yellow shaded part, there is $\Delta>|\eta_{0(1)}|$.} 
\label{s2}
\end{figure*}

\bigskip
\section{Optimization for the single-qubit gate}

In our model, the single-qubit gates are operated with the resonant
microwave driving pulse plus a very small offset which can be optimized in the simulation process, then there is $\omega_{d_{l}}=\tilde{\omega}_{l}+\delta \tilde{\omega}_{l}$ and $\delta \tilde{\omega}_{l}\ll \tilde{\omega}_l/10$. 

We use a shaped pulse, e.g., derivative reduction by adiabatic gate (DRAG) pulse \cite{motzoi2009simple,chen2016measuring}, to implement our single qubit gate with 

\begin{equation}
\Omega_{X_{l}}(t)=A_{l}[1-\cos(\frac{2\pi}{T_g}t)]
\end{equation}

and 
\begin{equation}
\Omega_{Y_{l}}(t)=-\frac{\alpha_l}{\eta_l}\dot{\Omega}_{X_l}(t),
\end{equation}
where $A_l$ is the optimized driving amplitude, $T_g$ is the gate time, and $\delta \tilde{\omega}_{l}$ and $\alpha_l$ are also the optimized parameters.

\bigskip
\section{Numerical example}

In the main text, we gave the numerical example for the case of the $\eta_l/2\pi=-0.3\,\rm GHz$ ($l=0,1$). To verify the generality, we have also calculated the case with  $\eta_l/2\pi=-0.2\,\rm GHz$, as shown in FIG. \ref{s2}, which shows the similar results as the one with $\eta_l/2\pi=-0.3\,\rm GHz$ in the main text.

\bigskip

\section{Error mitigation for two-qubit gate}
For two-qubit gate, there is one case, in which we can apply our crosstalk-error mitigation method for single-qubit gate directly. When operating the two-qubit gate, there will be the additional phase gate to compensate the local phase of single qubit \cite{zhao2020switchable}, like $U(\beta_0,\beta_1) = e^{-i\beta_0 ZI/2}e^{-i\beta_1 IZ/2}$, where $\beta_{0(1)}$ can be optimized to minimize the infidelity of the two-qubit gate. When operating the simultaneous two-qubit gates, the local phase which is optimized with the isolated tow-qubit gate will change. This kind of error also belongs to the error constrained in the single-qubit computational space, so that, we can corrected it by optimizing the parameters in the phase gate $U(\beta_0,\beta_1)$.

\end{document}